# Benchmarking the viability of 3D printed micromodels for single phase flow using Particle Image Velocimetry and Direct Numerical Simulations

Alexandros Patsoukis Dimou · Hannah P. Menke · Julien Maes

## Abstract

Holistic understanding of multiphase reactive flow mechanisms such as $CO_2$ dissolution, multiphase displacement, and snap-off events are vital for optimisation of large-scale industrial operations like $CO_2$ sequestration, enhanced oil recovery, and geothermal energy. Recent advances in three-dimensional (3D) printing allow for cheap and fast manufacturing of complex porosity models, which enable investigation of specific flow processes in a repeatable manner as well as sensitivity analysis for small geometry alterations. However, there are concerns regarding dimensional fidelity, shape conformity and surface quality, and therefore the printing quality and printer limitations must be benchmarked.

We present an experimental investigation into the ability of 3D printing to generate custom-designed micromodels accurately and repeatably down to a minimum pore throat size of 140 μm, which is representative of the average pore-throat size in coarse sandstones. Homogeneous and heterogeneous micromodel geometries are designed, then the 3D printing process is optimised to achieve repeatable experiments with single-phase fluid flow. Finally, Particle Image Velocimetry is used to compare the velocity map obtained from flow experiments in 3D printed micromodels with the map generated with direct numerical simulation (OpenFOAM software) and an accurate match is obtained. This work indicates that 3D printed micromodels can be used to accurately investigate pore-scale processes present in $CO_2$ sequestration, enhanced oil recovery and geothermal energy applications more cheaply than traditional micromodel methods.

## 1.Introduction

Sustainable low-carbon energy production is one of the major challenges society faces today. The intergovernmental Panel on Climate Change (IPCC) has stated that negative emissions via carbon capture and storage (CCS) are vital to mitigate the effects of global climate (Metz et al., 2005) (Committee on Climate Change, 2019). Improving engineering of the subsurface for oil and gas production, low-carbon energy storage, and $CO_2$ trapping is a crucial aspect of lowering carbon emissions.

An in-depth understanding of flow in porous media is critical for control and optimisation of these processes. However, the various mechanisms controlling the movement of fluids in the pore space (e.g. viscous displacement, capillary driven flow, and spontaneous imbibition) occur at the pore-scale and are poorly characterised (Blunt, 2017). Multiphase fluid displacement is an important process during enhanced oil recovery (Szulczewski, 2012) and $CO_2$ sequestration (Blunt et al., 2013, Orr Fm Jr, 1984). Fluids can be displaced in different ways depending on the physical and chemical properties of the two fluids as well as structural and surface properties of the medium itself (Zhang, 2011). Yet there is still little understanding on how those structural and surface properties impact the dynamic multiphase fluid arrangements, which makes the optimization and upscaling of these processes for continuum-scale prediction challenging.

Alexandros Patsoukis Dimou
ap92@hw.ac.uk
Institute of Geoenergy and Engineering, EH14 4AS Heriot Watt University, United Kingdom

X-Ray Computed Micro- and Nano-Tomography (X-Ray CT) has made it possible to image, observe, and quantify the 3D physio-chemical interactions between fluids and rocks at the scale of pores and grains. X-Ray CT imaging, in combination with state-of-the-art numerical simulations (Raeini et al., 2019); (Faris et al., 2020), has considerably improved our understanding of the fundamental physics that govern the fluid-fluid-rock interactions, and yet there are still several major challenges. First, the repeatability of experiments and second, the validation of numerical models. Even in relatively homogeneous reservoirs, each rock sample is unique and its properties unknown *a priori* (e.g. wettability, surface roughness). Furthermore, flow experiments can irreparably alter rock properties such as roughness and wettability. It is therefore extremely challenging to conduct fully controlled and repeatable experiments in a real rock sample. Additionally, X-Ray CT has low image acquisition speeds that are orders of magnitude higher than the timescale of pore filling events such as Haines jumps and thus the dynamics can be missed. Reproducibility of the experiments and time acquisition constraints render the validation of the numerical simulations difficult and validation is essential since different numerical methods may produce extensively different results (Zhao et al., 2019).

Micromodels are simplified, two-dimensional porous media that have standardised, repeatable geometries (Buchgraber et al., 2012) and have contributed significantly to our understanding of pore-scale physics and transport (Sun et al., 2016). Micromodel experiments can be conducted in the same geometry with multiple experimental protocols. Furthermore, they allow for geometry control and precise interrogation of the impact of structure on flow and transport. Quantifying the interplay between structure and pore-scale flow phenomena will allow for upscaling, optimising and predicting flow behaviour at the reservoir scale. In addition, due to their transparent nature, micromodels permit direct fluid flow visualisation with a high-resolution camera. The output data can be therefore augmented using techniques like Particle Image Velocimetry (PIV), which is a non-intrusive analysis technique where the particle distribution inside the domain is recorded at two instances of time. Using this change in particle distribution over time we are able to map the velocity field (Roman et al., 2016, Lindken et al., 2009). However, conventional micromodel fabrication techniques like micromodel etching, and moulding are expensive, slow and so far limited to 2D structures.

The emergence of additive manufacturing, also called 3D printing, offers a compelling alternative to conventional micromodels. 3D printing converts computer assisted design (CAD) into a physical object in a single process. Commercial 3D printers, which are capable of producing structures ranging from few microns to several centimetres, are beginning to challenge soft lithography as the research prototyping approach to micro-fabrication (Waheed et al., 2016). 3D printing has been applied to a wide range of industries including medicine (Vukicevic et al., 2017), biomedical engineering (Beg et al., 2020) and aerospace engineering (Joshi and Sheikh, 2015). In comparison with standard micromodel fabrication techniques, the attraction of 3D printing is twofold. First, 3D printing has an unprecedented potential to fabricate in three dimensions in a way that has not been previously possible. Second, the inexpensive nature of the 3D printed micromodels combined with the fast fabrication (~3 hours) allows experimental investigations on multiple geometries as well as to generating and quick testing to identify the optimal geometry that will produce the experimental data required. (Waheed et al., 2016). Small alterations of 3D printed models also enable geometrical sensitivity analysis, something that is not commercially possible with the existing fabrication techniques like micromodel etching and moulding. However, there are concerns regarding dimensional fidelity, shape conformity, surface quality (Waheed et al., 2016). Ahkami et al. (2019) observed irregular square pillar print an non-homogeneous dimensions of the pillars when they attempted to generate 3D printed micromodels. Watson et al. (2019) observed irregular channel width. Furthermore, Suzuki et al. (2016) observed blocked fractures when generating a 3-dimensional fracture network, which could compromise the integrity of experimental results. Therefore, the printing quality and printers' limitations must be investigated to verify their suitability for studying pore-scale phenomena and that is the aim of this study.

In this paper we have three objectives: (1) to investigate whether the single-phase flow PIV technique can be applied with our experimental setup; (2) to confirm the ability of our 3D printer to generate repeatable and accurate homogeneous micromodels at realistic pore sizes; And finally (3) test the limitations of our 3D printer in generating heterogeneous micromodels with realistic pore size distributions. Validation of the geometry of the 3D printed products will include comparison between the PIV results and direct numerical simulation using the OpenFOAM, the opensource Computational Fluid Dynamics (CFD) platform (OPENCFD, 2016).

## 2.Methods

### 2.1 Micromodel Geometries

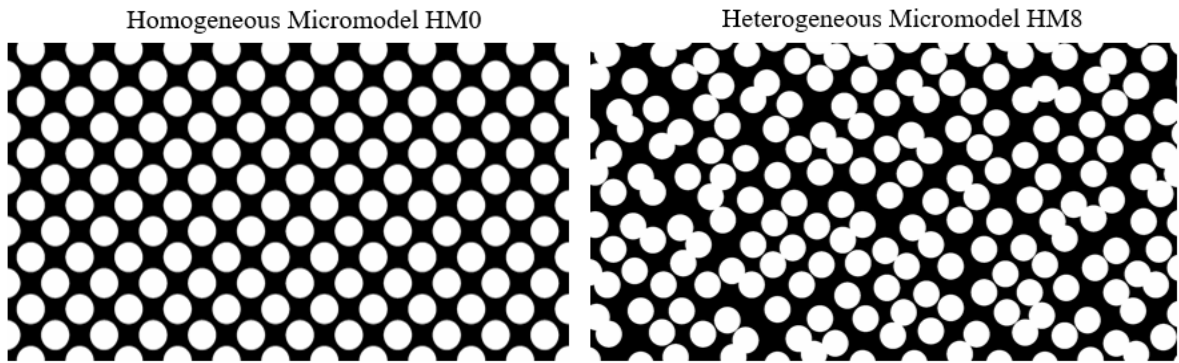

*Figure 1 : Generated micromodel geometries. Homogeneous micromodel HM0 (left) and Heterogeneous micromodel HM8 (right).*

The micromodel bead spacing was designed using python using the matplotlib (Hunter, 2007) and numpy libraries (Harris et al., 2020). A program was written to draw filled circles (grains) on an evenly spaced diagonal grid with a random deviation. In this way we were able to stochastically create infinite micromodel designs, all with the same overall statistical pore-space heterogeneity. The maximum allowable deviation can then be controlled. HM0 has no deviation, while HM8 has random deviation between zero and the whole average distance between each grain, which allows grains to touch but not to overlap (Figure 1).

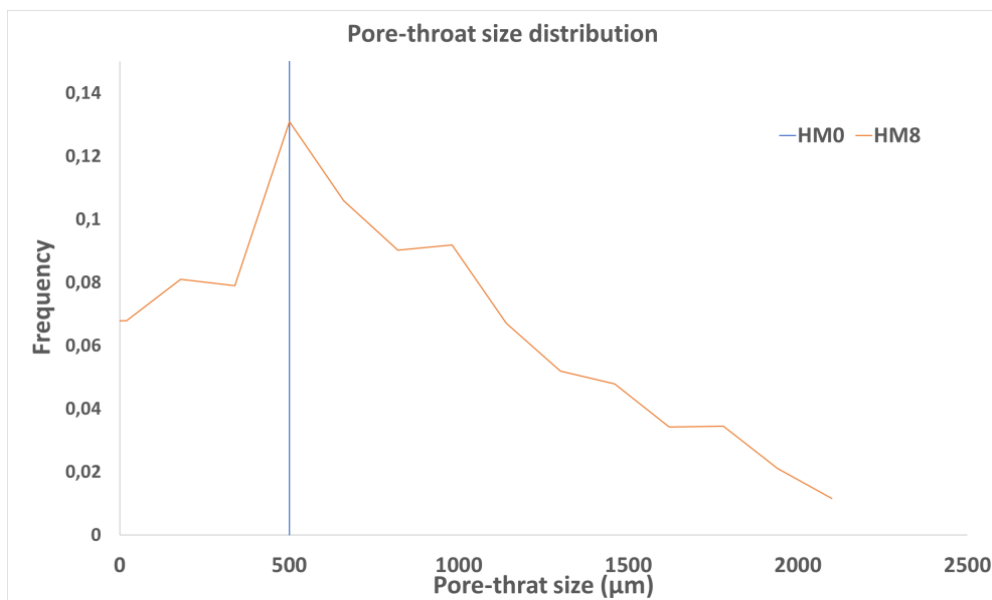

*Figure 2: Pore-size distributions of generated micromodels HM0 and HM8.*

## 2.2 Micromodel Fabrication

The micromodel designs were then transformed to 3D geometries using FreeCad software (Figure 3B) https://www.freecadweb.org/. The pattern was printed on a 25x25mm disk size in order to fit inside the visualization cell and the pattern depth was set to 200 μm. After the micromodel geometry was rendered as stereolithography file (STL) it was uploaded to the 3D printer (Figure 3C). In order to quantify the ability of the printer to generate repeatable micromodels, three micromodels of each geometry were printed for each pattern.

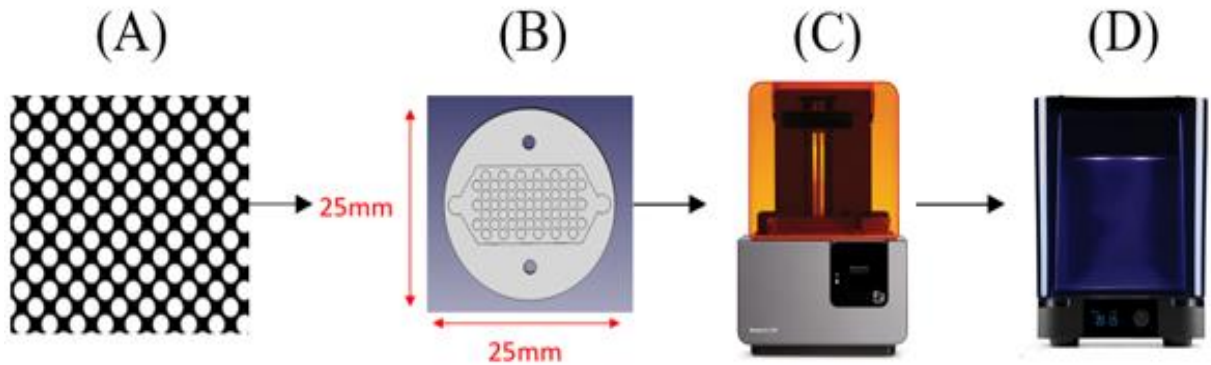

*Figure 3: Micromodel fabrication process. After the binary image is constructed (A) it is transformed to a 3D object (B). Then the 3D object is printed with the Formlabs Form 2 printer (C). After the 3D printed process is printed the micromodels are washed in IPA with the Formlabs Form Wash to clean from unsolidified resin (D).*

*Table 1: Printer resolution for different axis*

| Axis | Resolution (μm) |
|:---:|:---:|
| X | 140 |
| Y | 140 |
| Z | 25-100 |

The printer used for the creation of the micromodel in this project is Formlabs Form 2 stereo lithography apparatus (SLA) printer (Figure 3C) (https://formlabs.com). Form 2 works by successively printing layers of material one on the top of the other. Printing is controlled by photo-polymerisation of a liquid resin. The resin is hardened in the desired shape by a scanning laser. The printer uses the constrained surface approach or 'bat' configuration. In the bat configuration, the movable substrate is suspended above the resin reservoir. The laser is below the tank, which has a transparent bottom. The resolution of the printer can be seen as presented in the manual can be seen in Table 1. After the micromodels are printed, they are inserted in the Formlabs Form Wash machine for 10 minutes, which is an Isopropyl Alcohol (IPA) washer that removes the uncross linked resin that covers the micromodel during the printing process to reveal the intended printed geometry (Figure 3D).

During the printing process the printer offers the option of printing with z axis resolution of 25 to 100 micron (Table 1). These numbers refer to the thickness of the layer that will be solidified at every successive step. In order to test how thickness affects the final print product we printed the micromodels with both resolutions 25 and 100 micron respectively. For printing a 200μm depth pore-throat, 8 layers of 25 μm layers are required while 2 layers of 100 μm are required. We found that in the case that 25 μm resolution was selected, the 200μm pore-throat was blocked, while when 100 μm resolution was selected the very same pore-throat was printed successfully (Figure 4). The extra layers required when selecting the 25 μm resolution setting increased the probability of error. Decreasing the z resolution to 100 μm allowed us to print the same geometry but with fewer print layers, therefore reducing the probability of error.

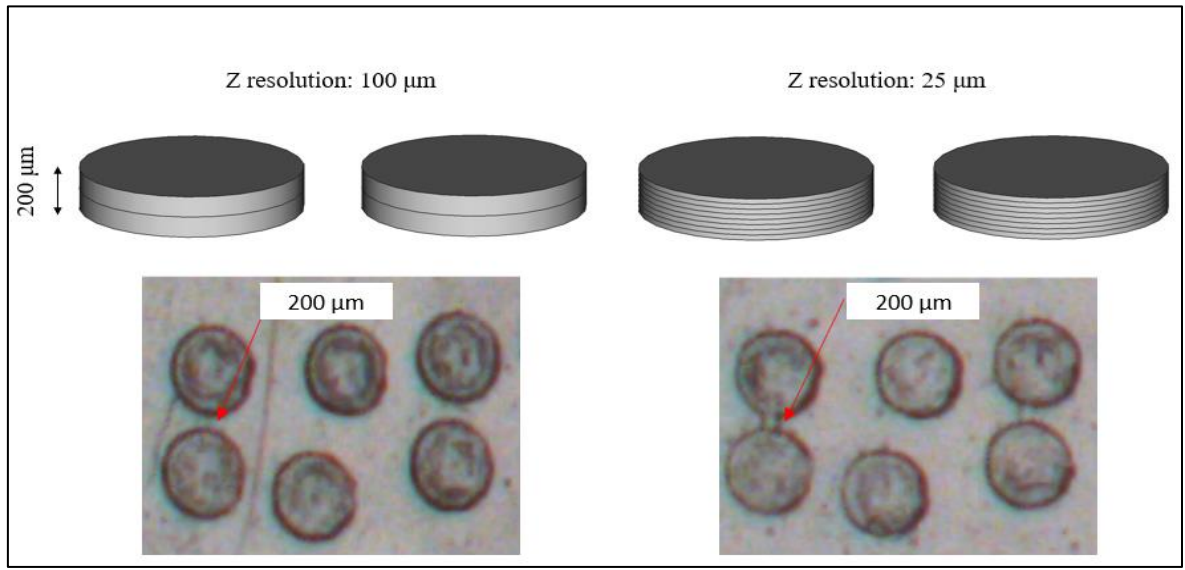

*Figure 4: Effect of Z resolution settings on printing quality*

To investigate whether the print geometry was correct after the micromodel was printed, its image was superimposed with the Free Cad design. We found that when printing at such small sizes there was a systematic error at the size of the pillar's diameter printed of the order of 80 micron (Figure 5). By taking into account the error during the design process (i.e. by designing the pillars with a diameter 80 micron smaller and by keeping their centre in the same position) we observed that the pillars were then printed at the intended diameter.

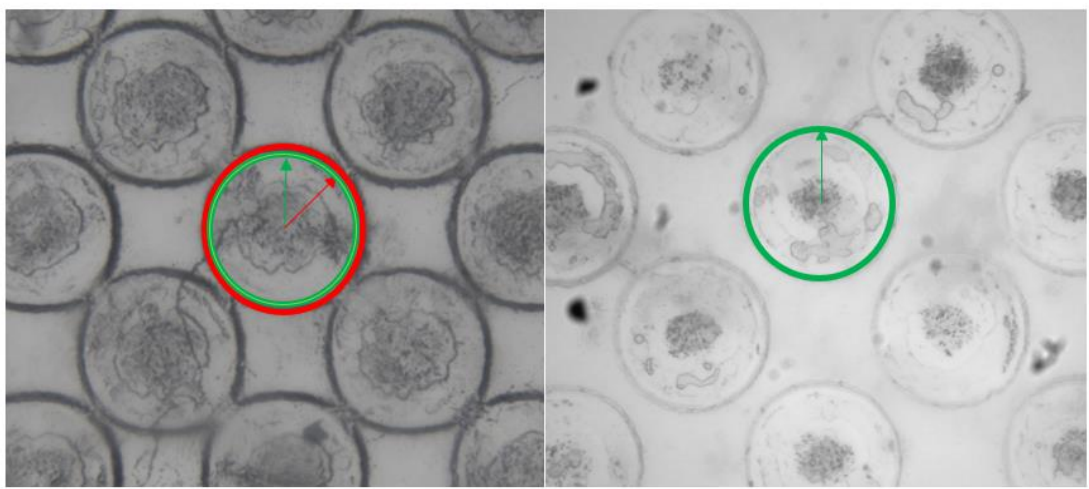

*Figure 5: Printing error on the bead radius. (Left) Print without taking account for error during design process. (Right) Print when taking print error during designing process. Intended radius (green) printed radius with an error of 80 μm (red).*

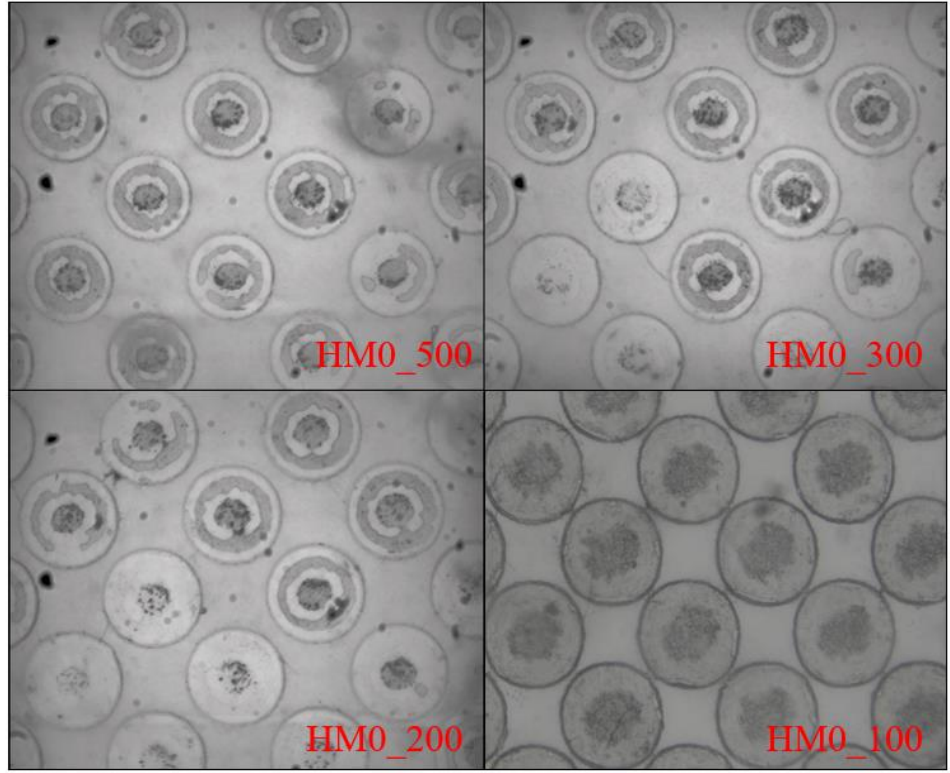

*Figure 6: Homogeneous micromodel final product image with high resolution camera.*

The 3D printer by following the process described above can successfully generate the homogeneous HM0 micromodels without blockage down to pore-throat radius of 200 μm. The micromodel HM0_100 which has pore-throat radiuses of 100 μm cannot be generated repeatably without blockage (Figure 6).

## 2.3 Experimental Setup

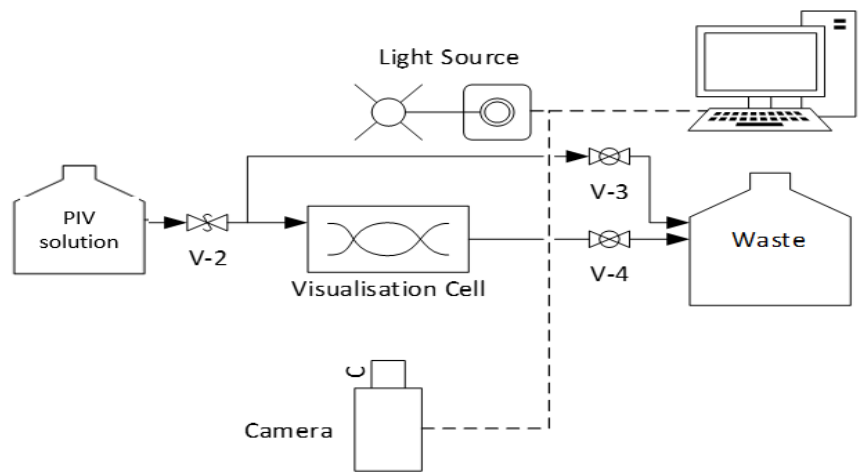

*Figure 7: Experimental setup including syringe pump (Chemyx Fusion 400) for PIV solution injection, visualisation perspex cell for sealing micromodel, high resolution camera (Baumer VCXU 51) and light source (SCHOTT ColdVision Light Source).*

Once the micromodel was successfully printed; it was inserted in the perspex transparent visualisation cell face-down and sealed using an o-ring. 1/16-inch peek tubing was used to connect the syringe pump (Chemyx Fusion 4000) to the visualisation cell and then to the outlet. A Baumer VCXU 51 high-resolution camera which allows for a 3.45 µm/pixel resolution was mounted beneath the flow cell and recorded images at 10 frame per second using the Stream Pix 11 software (https://www.norpix.com/products/streampix/streampix.php). Above the visualisation cell, a LED light source (SCHOTT ColdVision Light Source) was installed to reduce shadows and enable clear visualisation of the movement of the particles (Figure 6). For each experiment, deionised (DI) water was seeded with Carboxylate Modified Latex (CML) micro particles (Polybead® Microspheres 18328-5) with a diameter of 15 µm at a concentration of 0.06% w/v. The density of the particles was 1.05g.cm$^{-3}$, which was close to the water density and therefore minimizing sedimentation. The polybead solution was then pumped through the visualisation cell at constant flow rate and the images recorded.

The Reynolds number is the ratio of inertial forces to viscous forces and it is defined as:

$$Re = \frac{\rho v d}{\mu} \qquad (1)$$

Where ρ is the density, u the fluid velocity d the diameter of the flow pattern and µ the fluid viscosity. The flowrate was set to 0.05 ml.min$^{-1}$ in order to achieve Re = 0.0047 and a creeping flow regime. The temperature of the experiment was ambient temperature and the outlet pressure was atmospheric.

## 2.4 Image Processing

Fluid velocity was calculated by dividing the average length of particle displacements by the time between subsequent images. To enable accurate measurement of the particle displacement, the time interval between two images must be adequately long to have particle move several pixels but at the same time short enough to avoid excessive deformation of the pattern formed by particles from one image to the next one (Roman et al., 2016). The timespan between images was chosen so that the maximum displacement of particles from one frame to the other was about 3 particle diameters (45µm). An acquisition speed of 10 frames.s$^{-1}$ was adequate to achieve this particle displacement between subsequent images. Since the lens used in the camera assembly has a focal depth such that the movement of the particles is captured at many planes within the thickness (200 µm) of the micromodel, the micromodel was thus treated as 3-dimensional and the velocity map calculated using PIV corresponds to the average velocity of the different planes.

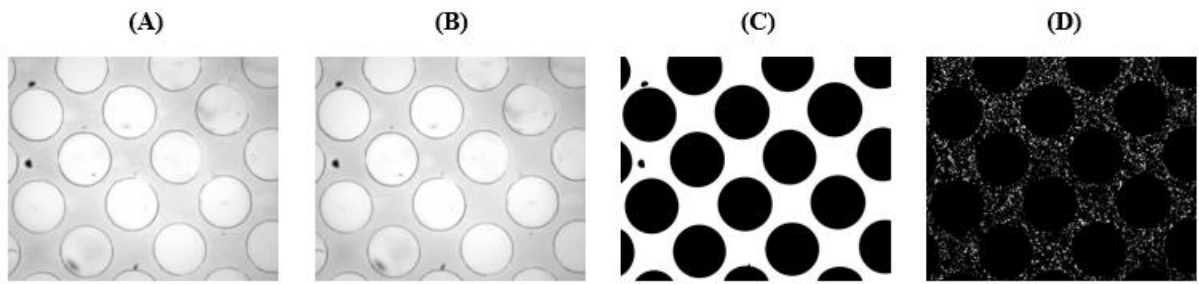

*Figure 8: Image processing for PIV analysis. Raw recorded Image (A). Image after background subtraction (B). Segmentation of the geometry to solid and flow- path (C). Final image where only the moving particles are visible (D).*

Images were pre-processed in MatLab (MATLAB, 2018) to improve the measurement quality. The micro-PIV measurements were done with PIVlab a MATLAB® tool (Thielicke and Stamhuis, 2014a). This tool allows image pre-processing before the images are correlated (Thielicke and Stamhuis, 2014b). First the images were denoising using the contrast limited adaptive histogram equalization

(Karel, 1994) and the adaptive wiener denoise filter (Lim, 1990). Uneven background illumination and noise was corrected by differential subtraction of a reference image from the PIV sequence. The pore-space was then segmented using intensity thresholding and used as a mask for PIV calculations. This pre- treatment allowed us to obtain sequences of images that contain only information regarding particle displacement (Roman et al., 2016). In order to cross-correlate the image data, PIV lab performed a direct Fourier transform correlation with multiple passes using the deforming windows algorithm (Roman et al., 2016) and velocity map with a resolution of 152x127 points grid was generated as output. Post-processing was performed by filtering outlier data and applying a physical local median filter.

## 2.5 Direct Numerical Simulation

To validate the velocity measurements from micro-PIV, we compared the experimental data with direct numerical simulation performed on the same geometries.

Flow in our system can be described by the Navier–Stokes equation which assuming steady-state and neglecting the effect of gravity is:

$$\mu \nabla^2 u = \rho(u \cdot \nabla)u + \nabla P \qquad (2)$$

where $\rho$ (kg.m$^{-3}$) is fluid density, u (m.sec$^{-1}$) is velocity, t (s) is time, $\mu$ (Pa.s) is dynamic viscosity, P (Pa) is pressure. And for incompressible flow we have the continuity equation:

$$\nabla \cdot u = 0 \qquad (3)$$

Numerical simulations were carried out using GeochemFoam ([www.julienmaes.com/geochemfoam](www.julienmaes.com/geochemfoam)), which is based on OpenFOAM, an opensource C$_{++}$ library designed to perform computational fluid dynamic calculations (OpenFoam 2017). A constant flow rate Q (m$^3$.sec$^{-1}$) with uniform pressure boundary conditions was applied at the inlet, with a non-slip boundary condition at the solid walls and free flow condition with constant pressure at the outlet. The flowrate selected was Q = 0.05 ml.min$^{-1}$, which is equal to that imposed during the experiments. To mesh the computational domain, a 3D uniform cartesian grid was first generated, and then cells containing solid were removed and replaced by cartesian cells to match the solid boundaries using the OpenFoam snappyHexMesh utility. The Grid selection was such that it matched the point resolution achieved with the PIV measurements and therefore a grid-block size of 55 x 55μm was selected (Figure 9).

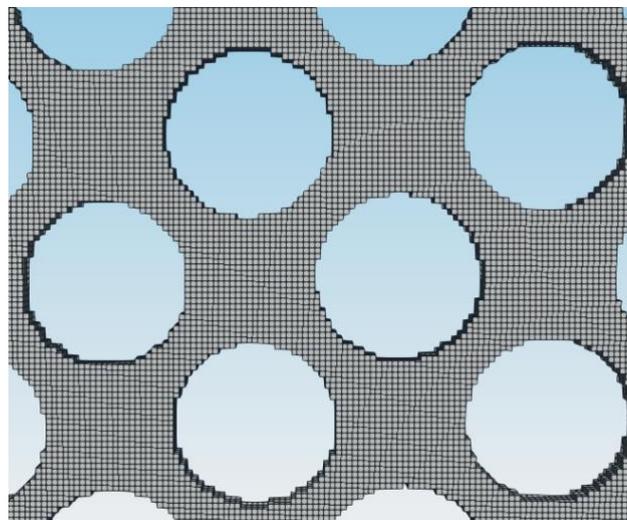

*Figure 9: Computational domain generated with OpenFoam. Blue circles represent solid bead. Grey boxes represent the mesh/computational domain.*

# 3. Results and Discussion

## 3.1 Homogeneous Micromodels

*Table 2 : L1 error when comparing PIV generated velocity maps and direct numerical simulation velocity maps for Homogeneous micromodels HM0 at different pore-throat sizes*

| Experiment | HM0_500(I) | HM0_500(II) | HM0_500(III) | HM0_500(IV) | HM0_500(V) |
|---|---|---|---|---|---|
| L1 error | 0.00871 | 0.0127 | 0.0131 | 0.00976 | 0.0102 |
| **Experiment** | **HM0_300(I)** | **HM0_300(II)** | **HM0_300(III)** | **HM0_300(IV)** | **HM0_300(V)** |
| L1 error | 0.0130 | 0.0129 | 0.00950 | 0.00932 | 0.0119 |
| **Experiment** | **HM0_200(I)** | **HM0_200(II)** | **HM0_200(III)** | **HM0_200(IV)** | **HM0_200(V)** |
| L1 error | 0.0476 | 0.011 | 0.0450 | 0.0128 | 0.0147 |

The error between the velocity map created by the PIV analysis and the DNS was calculated by:

$$L1\ Error = \frac{\frac{\sum_{i=1}^{Nu.Blocks}(U_{DNS(i)} - U_{PIV(i)})}{Nu.Blocks}}{Average\ Velocity} \qquad (4)$$

$U_{DNS}$ and $U_{PIV}$ refer to the velocity magnitude (m.sec$^{-1}$) calculated with DNS and PIV analysis respectively and subscript refers to the block inside the velocity map.

HM0 homogeneous micromodel was printed at 3 different pore-throat sizes (Table 1) to investigate the minimum pore-throat that can accurately be printed using our 3D printer. The micromodels were printed with pore throat sizes 500μm (HM0_500), 300μm (HM0_300) and 200μm (HM0_200). In addition, every micromodel size was printed five times and the experiment was repeated in every geometry in order to investigate the repeatability of each generated geometry.

With the use of MATLAB software the experimental velocity maps were then subtracted from the OpenFOAM generated velocity maps in order to obtain error maps for each experiment (Figure 10). From the resulting error maps Eq.4 was used to calculate the L1 error, the results of which can be found in table 2.

There was good agreement between the numerical simulation generated velocity maps and the velocity maps derived from PIV experiments conducted on the homogeneous micromodel HM0 at pore throat sizes 500 and 300 μm (HM0_500 & HM0_300). The L1 error calculated in all 5 experiments at each pore-throat size is less than 0.0131. The fact that the L1 error does not increase when the pore-throat radius of the micromodels is reduced from 500 μm to 300 μm suggests that the error present is not printing error but most probably an error related to the PIV acquisition. Additionally, when the pore-throat sizes of the HM0 micromodel is reduced to 200 μm (HM0_200), only three of the five geometries HM0_200(II), HM0_200(IV), HM_200(IV) have such a low error while the rest HM0_200(I) and HM0_200(III) show an increased error with an L1 of approximately 0.04, which suggest 200 μm is the pore-throat size close to the limit which our 3D printer can successfully generate pore-throat geometries. Finally, when the micromodel is designed to have 100 μm pore throat sizes blockage is always observed, and no PIV experiments could be attempted. This shows 3D printed micromodels can be successfully generated with negligible geometry error down to a pore-throat size of 300 μm while small error manifests when attempting to print 200 μm pore-throats. 100 μm pore-throats cannot be printed, while between 200 and 100 μm pore-throats is the limit of the printer to generate successfully and repeatably the required geometry.

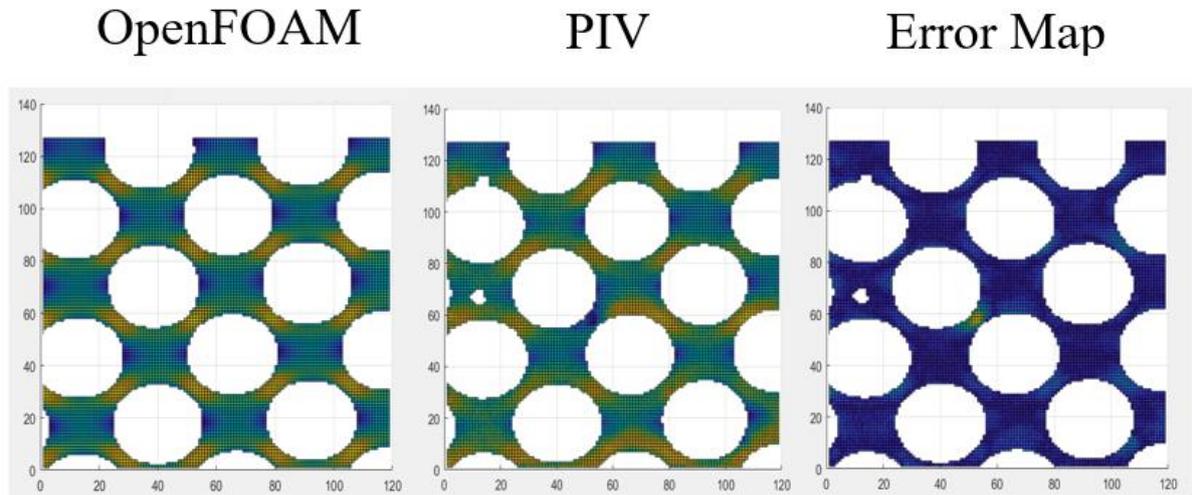

*Figure 10: PIV generated velocity map subtracted from OpenFOAM generated velocity map to calculate the error map which is used to quantify the experimental printing error.*

## 3.2 Heterogeneous Micromodels

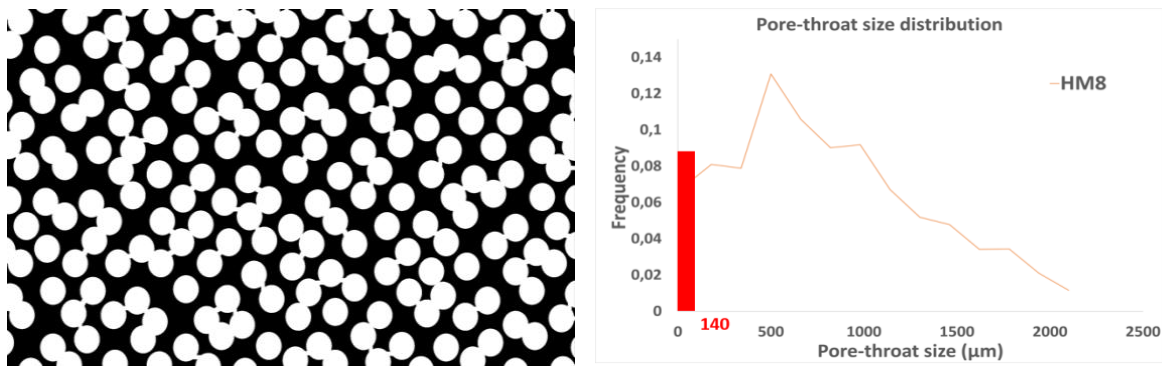

*Figure 11: HM8 micromodel after intentionally blocking all pore throats below 140 μm size. Pore size distribution of HM8 micromodel after intentional blockage.*

In order to fully capture a representative elementary volume (REV) of a heterogenous structure inside a micromodel it is necessary to have a large number of pores and throats. This is challenging in small model domains such as ours where we are limited by the printer resolution and the total maximum domain size (25x25 mm). These design limitations result in some of the throats being very close or below the printer's resolution limit and can result in throats below a threshold value being non-repeatably blocked in supposedly identical micromodel prints. Through trial and error that threshold pore throat diameter was identified to be 140μm. A MATLAB code was then written to close all the pore throats below 140μm such that they are intentionally blocked (Figure 11), and thus ensuring micromodel print repeatability. The effect of the intentional blockage of the throats on the throat size distribution of the pattern can be seen in Figure 11, where we can see that this truncation has very little effect on the overall distribution and closes less than 10% of the pore throats.

In table 3 the L1 error of the repeatably intentionally blocked micromodel is presented. Comparing the numerical simulation velocity maps generated with the 5 generated PIV velocity maps for the intentionally blocked HM8 micromodel (Figure 12) we get L1 error which is less than 0.0198. Therefore, the smallest size that the printer is capable of generating is 140 μm. It can also be seen that the printing error that manifests below 200 μm is not visible in the heterogeneous micromodel case since the flow occurring through the smallest pore-throats is less than in the larger pore-throats

*Table 3: L1 for Heterogeneous micromodel experiments HM8.*

| Experiment | HM8(I) | HM8(II) | HM8(III) | HM8(IV) | HM8(V) |
|---|---|---|---|---|---|
| L1 | 0.0198 | 0.0148 | 0.0096 | 0.0127 | 0.0093 |

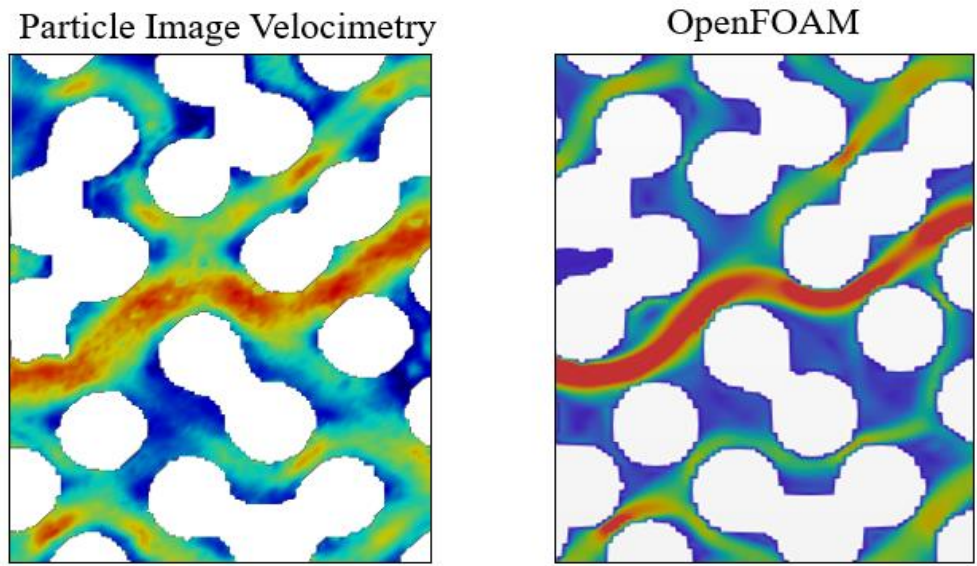

*Figure 12: Velocity maps for intentionally blocked Heterogeneous micromodel HM8. Solid beads are represented with white colour. Jet colourmap representing the velocity in the flow-path domain.*

## 4. CONCLUSIONS

In this paper we have investigated and showed that single-phase flow PIV works with our experimental setup. We show the ability of our 3D printer to generate repeatable homogeneous and heterogeneous micromodels at realistic pore-throat sizes by comparing single-phase flow PIV with direct numerical simulation results. The result obtained when comparing the numerical simulation generated velocity maps with the velocity maps produced from PIV experiments conducted on the homogeneous micromodel HM0 at pore throat sizes 500 and 300 μm show good agreement. This means that our 3D printer can successfully generate the required geometry every time. We have also shown that when the pore-throat sizes of the HM0 micromodel is reduced to 200 μm that only three of the five geometries have such a low error while the rest show an increased error with an L1 of approximately 0.04. Furthermore, when the micromodel is designed to have 100 μm pore throat sizes blockage is always observed. This suggests that our printer starts to struggle to generate the required size pore-throat sizes between 100 μm and 200 μm.

In the heterogeneous micromodel with some pore-throat sizes below 100 μm, we identified that non-repeatable blockage manifests for pore-throats with sizes < 140μm. Therefore, we can conclude that the limit of our 3D printer to print unblocked pore-throats is 140 μm. In order to generate repeatable heterogeneous micromodels pore-throats with sizes < 140 have to be intentionally blocked. Comparing the numerical simulation velocity maps generated with the five PIV velocity maps for the intentionally blocked HM8 micromodel, a very good agreement is observed. This indicates the 3D printed micromodels with minimum pore-throat size of 140 μm can be generated repeatably for one phase-flow experiments. This work proves that 3D printed micromodels with a specified geometry and a realistic pore size distribution can be repeatably and accurately generated and therefore there is potential to be used in the future for two-phase flow pore-scale investigations that manifest during applications like CCS, improved oil recovery and geothermal energy.

# Declarations

## Funding

## Conflict of interests/ Competing Interests
The authors declare no competing interests

## Author Contributions
A.P.D, H.P.M and J.M designed and performed this research. A.P.D wrote the manuscript. J.M and H.P.M reviewed the manuscript.

## Availability of data and material
Not applicable

## Code availability
Not applicable

multiphase flow in porous media. *Proceedings of the National Academy of Sciences of the United States of America,* 116**,** 13799-13806.